\documentstyle[prl,aps]{revtex}
\large
\begin{document}
\twocolumn

\draft

\title{A.C.-Transport Through a Two-Dimensional \\
Quantum Point Contact}

\author{{\bf I.E. Aronov$^{1,2}$, G.P. Berman$^{1,3}$, D.K. Campbell$^4$, and S.V. Dudiy$^2$}}

\address{
$^1$ Theoretical Division, MS-B213, Los Alamos National Laboratory, Los Alamos, New Mexico, 87545, USA,\\
$^2$ Institute for Radiophysics and Electronics, 
National Acadamy of Sciences of Ukraine, 310085, Kharkov, Ukraine,\\
$^3$ Kirensky Institute of Physics, 660036, Krasnoyarsk, Russia,\\
$^4$Department of Physics, University of Illinois at Urbana-Champaign,
1110 West Green St., Urbana, IL 61801-3080, U.S.A.}

\maketitle

\begin{abstract} 
 
    We calculate the admittance of a two-dimensional quantum point contact (QPC) using a Boltzman-like kinetic equation derived for a partial Wigner distribution function in an effective potential. We show that this approach leads to the known stepwise behavior of the admittance as a function of the gate voltage. The emittance contains both a quantum inductance determined by the harmonic mean of the velocities for the propagating electron modes and a quantum capacitance determined by the reflected modes.
\newline
\renewcommand{\baselinestretch}{1.656}

\end{abstract}

        Recent technological progress in manufacturing nano-scale solid
state structures has made it possible to fabricate devices in
two-dimensional electronic systems (2DES) which operate in the quantum
ballistic regime. A particular system that has attracted much attention
is the ballistic quantum point contact (QPC) \cite{glaz1,1,3,5,7,10,11,13,15,16,b1,imry}. 

In this letter we investigate a.c.-electron transport through a QPC. Despite
detailed investigations, both experimental \cite{1,3,5,7} and
theoretical \cite{glaz1,10,11,13,15,16,b1,imry}, of the d.c. transport,
a.c. transport effects have so far not drawn comparable attention, but see recent important papers \cite{13,15,16,b1} (and \cite{rt,rt1,rt2,rt3} for the  kinetic response of the resonant tunnel junction).

Several experiments \cite{1,3,5,7} have
exhibited such quantum coherent phenomenon as quantization of
the d.c.-conductance versus the gate voltage (or the number of propagating
modes through the QPC). The theory of this phenomenon
\cite{glaz1,10,11,13,15,16,b1,imry} explains the
d.c.-conductance quantization as a consequence of the
adiabaticity of the electrons' motion 
through the QPC with
smooth boundaries.
In an adiabatic geometry (see Fig.1), which
is smooth on the scale of the Fermi wavelength, the longitudinal and
transverse motion of electrons can be separated in the Schr\"{o}dinger
equation \cite{glaz1}. In this case, the number of the transverse quantization
modes is an adiabatic invariant, and the transverse energy plays the role of the
potential energy for the one-dimensional longitudinal motion of each mode.
Depending on whether the total energy of a given electron state is greater
or less than the effective potential energy of each mode, the mode
is propagating or non-propagating (see Fig. 2).

Frequency dependent transport phenomena in QPCs are clearly of interest,
for the a.c. frequency introduces a time-scale that may reveal qualitatively new
effects.

The a.c.-transport has been considered by M.~B\"{u}ttiker {\it et al}.
\cite{13,15,16,b1}, who showed that it is described at low frequency $\omega$,
by an a.c.-{\it admittance}, $Y \equiv 1/Z=G-i\omega{\cal E}$
($Z$ is the {\it impedance}). $G$ is the {\it conductance} and ${\cal E}$
is the {\it emittance}, a term first introduced in this context by
M.~B\"{u}ttiker \cite{16}.
Christen and B\"uttiker \cite{b1} derived
the general expression for the electrochemical capacitance and for the displacement current, and established  the steplike behavior of the QPC emittance in synchronism with the conductance steps, and discussed the admittance for quantized Hall conductors. In paper \cite{13,15,16,b1}, the emittance was expressed in terms of the geometric capacitance, transmission probability, and the densities of states of the ``mesoscopic capacitor plates'. 

Below we develop a simple method for calculating
the transport characteristics for the QPC, which is based on the Wigner
distribution function (WDF) formalism.  
(The effectiveness of
the WDF approach to modeling of the mesoscopic devices was demonstrated
in \cite{f1}.)Our approach allows us to represent the emittance in terms of the capacitance and the inductance, which are expressed in the explicit form through the microscopic characteristics.
 Assuming adibaticity, we derive a
Boltzmann-like quantum kinetic equation for the {\it partial WDF} for the QPC.
This equation allows us to treat the 2DES in a QPC in terms of classical
trajectories for the effective 1D motion.
We calculate the a.c.-admittance of the
QPC, considering propagating and non-propagating (reflected) electron
modes and show that the real part of the admittance (the conductance) is
quantized versus of the gate voltage, consistent with earlier
calculations \cite{glaz1} based on the Landauer formula \cite{20}. 
We also show that the emittance ${\cal E}$ has a negative part (the quantum
inductance), which arises from the propagating electron modes and
whose value is determined by the harmonic mean of the electron velocities.
The non-propagating electron modes give a
positive contribution (the quantum capacitance) to ${\cal E}$.
 This contribution is determined by the
which depends on a geometrical form of the QPC, which
is controlled by the gate voltage.

To find the conductivity of a 2DES of a QPC (see
Fig. 1) taking account of both frequency dependence and spatial dispersion,
we shall use an approach based on the Wigner distribution function \cite{18,19},
$f_{\vec p}^W(\vec r)=\int d\vec r^\prime Tr\{\hat\rho
\exp[-\frac{i}{\hbar}(\vec p+\frac{e}{c}\vec A(\vec r))
\vec r^\prime]
\Psi^+(\vec r-\vec r^\prime/2)\Psi(\vec r+\vec r^\prime/2)\}.
$
Here $\hat\rho$ is the density matrix operator of the system and
obeys the quantum Liouville equation;
$\Psi^+(\vec r)$ ($\Psi(\vec r)$) is the Fermi operator creating (annihilating)
a particle at $\vec r$; and $\vec A$
is the vector-potential of the electromagnetic field. When the
characteristic scale of the spatial inhomogeneities exceeds both
the radius of interaction among the particles and the electron
de Broglie wavelength, the kinetic
equation for the WDF takes the form equivalent to the classical
kinetic equation \cite{19}. From the WDF, we can find the charge density
$\rho$ and the current density $\vec j$ \cite{18,19}.

In the adiabatic model \cite{glaz1,imry} of the QPC shown in Fig. 1,
one assumes the constriction to be smoothly
varying $d^\prime(x)\simeq d(x)/\tilde L\ll 1$, where $2\tilde L$ is a
characteristic adiabatic length scale
of the QPC. With this assumption, the variables in the Schr\"{o}dinger equation
can be separated, and the eigen-wave function can be written in the form:
$\psi_n(x,y)=\psi_n(x)\Phi_n[y,d(x)],$
where the transverse wave function $\Phi_n(y)=(1/\sqrt{d(x)})\sin\left\{{\pi
n[y+d(x)]}/{2d(x)}\right\} \theta[d^2(x)-y^2]$,
satisfies the boundary conditions
$\Phi_n(y)|_{y=\pm d(x)}=0$,
and $\theta(x)$ is the Heaviside-step function. The effective Hamiltonian for
the longitudinal wave function $\psi_n(x)$ is
$\hat H=-({\hbar^2}/{2m})\partial_{xx}^2+\varepsilon_n(x)+e\phi(x)$,
where $\phi(x)$ is the electric potential averaged with respect to
the transverse coordinate $y$.

Adibaticity and the quantization of the transverse motion imply
that the energy
$\varepsilon_n(x)$ in the Hamiltonian $\hat H$ has the form
$\varepsilon_n(x)={\pi^2n^2\hbar^2}/{8md^2(x)}$, with
the transverse quantum number $n$ an adiabatic integral of motion.
Thus the motion of electrons in the QPC can be viewed as
a set of $n$ effective one-dimensional electron systems.
Each effective electron system is influenced by
both the potential $\varepsilon_n(x)$ and
the self-consistent electric potential $\phi(x)$. To describe these
separate systems we introduce a {\it partial WDF} (PWDF) via
$$
f_n^W(x,p_x)=\int dx^\prime\exp\left(-{ip_xx^\prime}/{\hbar}\right)
Tr\hat\rho\eqno(1)
$$
$$
\Psi^+_n(x-x^\prime/2)\Psi_n(x+x^\prime/2).
$$
Then we can represent the WDF in the form
$$
f_{\vec p}^W(\vec r)=\sum_{n=1}^\infty
f_n^W(x,p_x)\int\limits_{-\infty}^\infty dy^\prime
\exp\left(-{ip_yy^\prime}/{\hbar}\right)\eqno(2)
$$
$$
\Phi_{n,x}(y-y^\prime/2)\Phi_{n,x}(y+y^\prime/2).
$$
The equation for the PWDF (1) can be derived using the Wigner
transformation \cite{18,19}:
$$
\frac{\partial f_n^W}{\partial t}+
v_x\frac{\partial f_n^W}{\partial x}+
\left[-\frac{\partial\varepsilon_n(x)}{\partial x}+eE(x)\right]
\frac{\partial f_n^W}{\partial p}=\hat I_n\left\{f_{\vec p}^W\right\},\eqno(3)
$$

where $p\equiv p_x$ and
$E(x)=-{\partial \phi(x)}/{\partial x}$. The equation (3) is derived in the semi-classical approximation, under the condition $k_FL\gg 1$.
In terms of the PWDF the nonequilibrium charge density and current density
can be defined as:
$\rho(x,y)=\sum_{n=1}^\infty\rho_n(x)\Phi_n^2(y)$,\quad
$j(x,y)$=$\sum_{n=1}^\infty j_n(x)\Phi_n^2(y)$,
where $\rho_n(x)$ and $j_n(x)$ are the partial charge and current
densities: $\rho_n(x)=({e}/{\pi\hbar})\int\limits_{-\infty}^\infty dp
\left[f_n^W(x,p)-f_n^{W(0)}(x,p)\right]$, and
$j_n(x)=({e}/{\pi\hbar m})\int\limits_{-\infty}^\infty dppf_n^W(x,p)$.
Here, $f_n^{W(0)}(x,p)$ is the equilibrium PWDF.
The kinetic equation (3) for the PWDF has the form as the
classical kinetic equation in the effective potential $\varepsilon_n(x)$.
Thus we can solve this equation by the method of characteristics,
{\it i.e.}, by using the classical trajectories.

 The collision integral in (3) includes
quantum transitions \cite{19} and intermixing of the
different electron modes (the different PWDF).
In this letter, we will approximate the collision integral by a
single momentum relaxation frequency, $\nu$, so that
$\hat I_n\left\{f_{\vec p}^W\right\}=-\nu\left[f_n^W(x,p)-f_n^{W(0)}\right]$,
where $f_n^{W(0)}$ is the equilibrium PWDF.
The equilibrium distribution function $f_n^{W(0)}$ within the
adiabatic approximation is given by:
$f_n^{W(0)}(x,p)=n_F\left\{[{p^2/2m+\varepsilon_n(x)-\mu}]/{T}\right\}$, where
$n_F$ is the Fermi function with the effective chemical potential
$\mu-\varepsilon_n(x)$, where $\mu$ is the equilibrium chemical potential
of the 2DEG. The characteristics of kinetic equation (3) are the
phase trajectories of a one-dimensional motion in the potential
$\varepsilon_n(x)$, which is determined from the integral of motion, {\it viz.}
the total energy: $\varepsilon={p^2}/{2m}+\varepsilon_n(x)=const$.
We consider a symmetrical QPC, $d(x)=d(-x)$. In this case the phase portrait is shown in Fig. 2. The separatrix lines (the
heavy lines) which pass through the hyperbolic point $p=0$, $x=0$,
divide the phase space into four regions, within which four sets of
phase trajectories exist.

The regions of propagating trajectories
($\varepsilon>\varepsilon_n(0)$) occupy the regions (see Fig. 2):
1) $\varepsilon>\varepsilon_n(0),\ p>0; $
2) $\varepsilon>\varepsilon_n(0),\ p<0$; the
regions of non-propagating (reflecting) trajectories are 
($\varepsilon<\varepsilon_n(0)$):
3) $\varepsilon<\varepsilon_n(0),\ x>0$; and
4) $\varepsilon<\varepsilon_n(0),\ x<0$.
Within each region, one can find the solution of the kinetic equation for
the PWDF and derive the general formula for the partial charge $\rho_n$
and the current densities $j_n$. Here we consider the most interesting case,
when the temperature is very low ($T\rightarrow 0$, $T\ll\mu$), so that
we have a clear separation between open channels
($\varepsilon_n(0)<\mu$) and closed channels ($\varepsilon_n(0)>\mu$).

Transport through a QPC in principle involves
nonlocal (integral) operators in that the charge and current densities
at a given point $x$ can influenced by the electric field within the
whole conductor.
Nonetheless, it is well known that the static
conductance is (to a good approximation)
specified by the potential difference (bias voltage)
between the right and left reservoirs and that the detailed electric potential
profile does not influence it significantly \cite{glaz1}.
This result was derived using the Landauer formula \cite{imry,20}
in which the conductance is determined by the matrix of the
transmission coefficients of the electron waves corresponding
to different propagating modes. We can readily show that this result
follows from our PWDF approach. In the ballistic regime, when $L\ll l$,
($2L$ is the distance between the reservoirs, $l$ is the mean free path) and
at $\omega$, $\nu\rightarrow 0$, we obtain for the open channels
$j_n=({2e^2}/{h})V$, with $V=\int\limits_{-L}^L dxE(x)$,
and for the closed channels
$j_n$=$0$. Hence for the static conductance we get the familiar
result \cite{glaz1}:
$G={I}/{V}={2e^2{\cal N}}/h$,
where ${\cal N}$ is the number of the open channels:
${\cal N}$=$\left[\{2k_Fd(0)\}/{\pi}\right]$;\quad $\hbar k_F$=$\sqrt{2m\mu}$.
Here the brackets $[\cdots]$ stand for an integer part. Clearly, the
static conductance does not depend on the specific form of the smooth function,
$d(x)$.

More generally, we can use the formalism of the PWDF to calculate
the admittance at a finite frequency $\omega$. The partial current $j_n$
is a function of the longitudinal coordinate $x$ at $\omega\ne 0$. 

The continuity equation, $div \vec j+{\partial\rho}/{\partial t}=0$,
in the QPC at $\omega\ne 0$ allows one to show that the total current,
$I_{tot} \equiv \sum_{n=1}^\infty\{j_n-i\omega
\int\limits_{-L}^{x}dx^\prime\rho_n(x^\prime)\}$,
.
which includes the electron current
$\sum_{n=1}^\infty j_n(x)$ and displacement current
$-i\omega\sum_{n=1}^\infty
\int\limits_{-L}^xdx^\prime\rho_n(x^\prime)$,
is independent of the longitudinal coordinate $x$. Within the left
reservoir the displacement current vanishes, so the total current is
$I_{tot}=\sum_{n=1}^\infty j_n(-L)$,
and the admittance can be determined as
$Y \equiv{I_{tot}}/{V}=({1}/{V})\sum_{n=1}^\infty j_n(-L)$.
In the general case, we should determine the field $E(x)$ self-consistently
within the QPC from
Maxwell's equations and afterwards calculate the admittance. For
the present, we
consider the long-wavelength approximation, where
$v_n^*\gg\omega L_n$.
Here $v_n^*$ is the typical velocity for electrons in the $n^{th}$
channel, and $L_n$ characterizes the length of a region for that channel. For
the propagating modes, $L_n$ is the distance between the reservoirs
($L_n\sim 2L$) and
$v_n^*=v_n(0)$.
For reflecting modes, $L_n$ is the doubled
distance between the turning points ($\varepsilon_n(x_n)=\mu$) and the nearest reservoir
($L_n\sim 2(L-x_n)$). Hence the typical velocity in this case is:
$v_n^*={2v_F}\sqrt{x_n(L-x_n)}/\tilde L$,  $v_F=\sqrt{{2\mu}/{m}}$,
where $2\tilde L$ is, as above, the characteristic adiabatic length scale of the
QPC.
To calculate the current of the propagating modes,
we can approximate the velocity $v_n$ as
${v_n(x)}\simeq{v_n(0)}={v_n^*}$,
while for the reflecting modes:
${v_n(x)}\simeq{v_n^*}\sqrt{{(|x|-x_n)}/{(L-x_n)}}$.
Consistent with the symmetry of $d(x)$ we assume that $E(x)=E(-x)$ within
the QPC and take $d(x)=d_0\exp\left[(x/\tilde L)^2\right]$.
The contribution of the propagating modes to the total current is
determined by the bias voltage $V$ and is independent of the detailed
profile of the electric potential inside the QPC, while that
of the reflecting modes depends on $d(x)$ and $E(x)$. We can write the admittance
in the form $Y=G-i\omega{\cal E}$,
where as shown above the static conductance is given by
$G=(2e^2/h){\cal N}$ and (after some calculation) one can show that the emittance
${\cal E}$ can be expressed as
$$
{\cal E}=-G\frac{L}{\overline{v}^{(o)}}+C,\quad C=
\frac{16}{3}\frac{e^2}{h}
\sum_{n={\cal N}+1}^{{\cal N}+\tilde{\cal N}}\frac{\xi_n}{v_n^*}(L-x_n).\eqno(4)
$$
Here $\overline{v}^{(o)}$ is the harmonic mean  of the velocities $v_n^*$
 in the propagating modes:
${1}/{\overline{v}^{(o)}}=(\frac{1}{{\cal N}})
\sum_{n=1}^{{\cal N}}{1}/{v_n^*}$.
The integer $\tilde{\cal N}$ determines the number of reflecting modes:
$\tilde{\cal N}=\left[({2k_Fd_0}/{\pi})\exp[(L/\tilde L)^2]
\right]-{\cal N}$,
$2L$ being the distance between the reservoirs. The dimensionless parameter
value $\xi_n$ reflects the form of the electric field
in the region $(x_n,L)$ filled with the electrons of the $n^{th}$ reflecting channel:
$\xi_n=({3}/{2})\int\limits_{x_n}^Ldx^\prime
({E(x\prime)}/{V})\sqrt{{(x^\prime-x_n)}/{(L-x_n)}}$.

From their respective expressions one sees immediately that
the contribution to ${\cal E}$ of the reflecting modes is {\it positive}
while that of the propagating modes is {\it negative}. In this situation we can
interpret our results in terms of an equivalent circuit in which a capacitance
and resistance are in parallel followed by an inductance in series. The
admittance of this circuit is
$$
Y= G-i\omega(C-{{\Lambda G^2}\over{c^2}}),\quad\omega C\ll G, \quad \omega\Lambda\ll {{c^2}\over{G}}.\eqno(5)
$$
The effective inductance in (5) is given by
$\Lambda={c^2L}/{G\bar v^{(o)}}$,
and the effective capacitance $C$ is given in (4).
Note, that the formula (5) coincides with the general expression for the emittance derived in \cite{b1} (see formula (7) in \cite{b1}), where the emittance was expressed in terms of the geometric capacitance, transmission probability, and the densities of states of the ``mesoscopic capacitor plates''.
Our approach allowed us to represent the emittance in terms of the capacitance and the inductance, which are represented in the explicit form through the microscopic characteristics.

The dependence of the emittance on the number of open channals
shows immediately that it is a stepwise function of the
gate voltage. 
When the gate voltage approaches a point for which $2k_Fd/\pi$
is an integer, so that an additional channel opens (or closes), our
expressions for the inductance and the capacitance diverge.
In this case, our assumption that $v_n^*\gg\omega L_n$ is violated, and
the contribution of these points into the admittance should be calculated separately,
using the full self-consistent Maxwell equation approach
 $\gamma_n=(\mu-\varepsilon_n(0))/\varepsilon_n(0)$,

In conclusion, we
have developed a new approach, based on a partial Wigner distribution function,
to analyze the a.c. electron transport properties of a quantum point
contact.
Treating the ballistic QPC in the adiabatic approximation,
we derived a Boltzman-like equation for the partial Wigner
distribution function in an effective potential brought about by the quantized
transverse modes. We analyzed this equation in terms
of propagating and reflecting trajectories in the
semi-classical approximation.

Our results establish that the a.c. electron transport depends
directly on the the number of propagating and reflecting
modes and that certain features are sensitive to
the form of the distribution of the
electric field in the QPC. In particular, the real part of the admittance (the
conductance) is determined by the number of propagating electron modes
and does not depend on the spatial distribution of the electric field
inside the QPC \cite{glaz1}.  The imaginary part of the admittance
(the emittance) exhibits stepwise behavior
as a function of the gate
voltage and consists of two parts: the quantum inductance and
the quantum capacitance. The quantum inductance is determined by the
harmonic mean of the velocities for the propagating electron modes. The
quantum capacitance is specified by the reflecting
modes and is sensitive to the geometry of the QPC.
The emittance can be controlled by the gate voltage

We stress that the effective quantum inductance
and capacitance, and the equivalent circuit, are concepts
valid within the linear response, low-frequency approximation.

For the high-frequency case, and when new propagating and non-propagating modes
can appear or disappear, the frequency dispersion of the admittance is more
complicated than the linear one given by the equivalent circuit
of Eq. (5). This case
must be considered using the self-consistent Maxwell equations for
the electric field in the QPC. We are presently investigating
this problem.

 We are grateful to L.I. Glazman, D.K. Ferry, R. Akis and
G.D. Doolen for fruitful discussions. This work was partially supported
by the Linkage Grant 93-1602 from the NATO Special Programme Panel
on Nanotechnology and by the INTAS Grant No. 94-3862.
%
      
%-----------------------------------------------------------------------

%---------------------------------------------------------------

%---------------------------------------------------------------
\newpage
\begin{center}
{\bf FIGURE CAPTIONS}\\ \ \\
\end{center}
\quad\\
Fig. 1: {The geometry of the microconstriction. The width is denoted by
$2d(x)$, the narrowest width is $2d_0$, and the effective length is
$2\tilde L$.}\\ \ \\
Fig. 2: {The plane of phase trajectories for one-dimensional motion
determined by conservation of integrals of motion. The heavy lines are
separatrices that separate the propagating modes (sections 1 and 2) and
non-propagating (reflecting) modes (sections 3 and 4)}.\\ \ \\
%

%---------------------------------------------------------------


\begin{references}

\bibitem{glaz1} L. I. Glazman, G. B. Lesovik, D. E. Khmelnitskii, R.Shekhter, {\it JETP
Lett.} {\bf 48}, 238 (1988).

\bibitem{1} C. W. J. Beenakker, H. van Hasten, p. 1 in  {\it Solid  State
Physics}, Vol. {\bf 44},  Eds: H.  Ehrenreich, D. Turnbull, (Academic, San Diego,
1991);
B.  J. van Wees, H. van Houten, C.  W. J. Beenakker, J. G.
Williamson, L. P. Kouwenhoven, D. van der Mare, C. T. Foxon, {\it Phys. Rev. Lett.}
{\bf 60}, 848  (1988).

\bibitem{3} D. A. Wharam, T. J.  Thornton, R.  Newbury, M. Pepper, H.
Ahmed, J. E. F. Frost, G. G. Hasko, D. C. Peacock, D. A. Ritchie, G.A. C. Jones,
{\it J. Phys. C} {\bf 21}, L209 (1988).

\bibitem{5} N. K. Patel, J. T.  Nicholls, L. Martin - Moreno, M. Pepper,
J. E. F. Frost, D. A. Ritchie, and G. A. C. Jones, {\it Phys. Rev. B} {\bf 44},
13549 (1991);
R. Taboryski, A. K. Geim, M. Persson, and P. E. Lindelof, {\it Phys. Rev. B}
{\bf 49}, 7813 (1994).

\bibitem{7} T. M. Eiles, J.  A. Simmons,  M. E. Sherwin,  and J. F. Klem,
{\it Phys. Rev. B} {\bf 52}, 10756 (1995).

\bibitem{10} I. E. Aronov,  M. Jonson, and A. M. Zagoskin, {\it Phys. Rev. B}
{\bf 50},  4590 (1994).

\bibitem{11}L. Gorelik, A. Grincwaig, V. Kleiner, R.Shekhter, M.
Jonson, {\it Phys. Rev. Lett.} {\bf 73}, 2260 (1994);
A. Grincwaig,  L. Gorelik, V.  Kleiner, and R.  I. Shekhter, {\it Phys. Rev. B}
{\bf 52}, 12168 (1995);
F. Hekking  and Yu. V. Nazarov, {\it Phys. Rev. B} {\bf 44}, 11506
(1991); {\it Phys. Rev. B} {\bf 44}, 9110 (1991).

\bibitem{13} M. B\"{u}ttiker, H. Thomas, and Pr\^{e}tre, {\it Phys. Lett. A}
{\bf 180}, 364 (1993);
M. B\"{u}ttiker, {\it J. Phys.: Condens. Matter} {\bf 5}, 9361 (1993).

\bibitem{15} M. B\"{u}ttiker, {\it Il Nuovo Cimento} {\bf 110B}, 509 (1995).

\bibitem{16} M. B\"{u}ttiker, In: {\it Quantum Dynamics of Submicron Structures},
Eds: H. A. Cerdeira {\it et al.}, p.p. 657-672 (1995, Kluwer Academic
Publishers, Netherlands).

\bibitem{b1} T. Christen, M. B\"{u}ttiker, {\it Phys. Rev. Lett.} {\bf 77}, 143 (1996);
T. Christen, M. B\"{u}ttiker, {\it Phys. Rev. B} {\bf 53}, 2064 (1996).

\bibitem{imry} A. Yacoby and Y. Imry, {\it Europhys. Lett.} {\bf 11}, 663 (1990).

\bibitem{rt} C.L. Fernando and W.R. Frensley, {\it Phys. Rev. B} {\bf 52}, 5092 (1995).

\bibitem{rt1} L.Y. Chen and C.S. Ting, {\it Phys. Rev. Lett.} {\bf 64}, 3159 (1990).

\bibitem{rt2} C. Jacoboni and P.Price, {\it Solid State Commun.} {\bf 75}, 193
(1990).
\bibitem{rt3} Y. Fu and S.C. Dudley, {\it Phys. Rev. Lett.} {\bf 70}, 3159 (1990);
C. Jacoboni and P.J. Price, {\it Phys. Rev. Lett.} {\bf 71}, 464 (1993);
Y. Fu and S.C. Dudley, {\it Phys. Rev. Lett.} {\bf 71}, 466 (1993).

\bibitem{f1}
D.K. Ferry and H.L. Grubin, {\it Modelling of Quantum Transport in
Semiconductor Devices}, p. 283, In: {\it Solid State Physics} {\bf 49} (1995);
N.C. Kluksdahl, A.M. Kriman, D.K. Ferry, C. Ringhofer,
{\it Phys. Rev. B} {\bf 39}, 7720 (1989).

\bibitem{20} R. Landauer, {\it IBM J. Res. Dev.} {\bf 1}, 233 (1957);
{\it Phil. Mag.} {\bf 21}, 863 (1970);{\it Z. Phys. B} {\bf 68}, 217 (1987).

\bibitem{18} E. Wigner, {\it Phys. Rev.} {\bf 40}, 749 (1932).

\bibitem{19} A. I. Akhiezer and S. V. Peletminskii, {\it Methods of
Statistical Physics}, (Oxford, New York, Pergamon Press, 1981).



\end{references}
\end{document}